\documentstyle[prl,twocolumn,aps]{revtex}
\begin{document}
\draft
\title
{Non-partial Reality}
\author{Yu Shi}
\address{Department of Physics, Bar-Ilan University, Ramat-Gan 52900, Israel}
\maketitle
\begin{abstract}
Study on pre- and postselected quantum system indicates that
``product rule'' and ``sum rule'' for elements of reality should be 
abandoned. We show that this so-called non-partial realism can refute 
 arguments  against hidden variables in a 
unified way, and might
save local realism. 
\end{abstract}
\pacs{PACS numbers: 03.65.Bz}
Einstein, Podolsky and Rosen (EPR) showed that  
quantum mechanical description of physical reality is 
not complete based on a sufficient condition for physical reality:
``If, without in any way disturbing a system, we
can predict with certainty (i.e., with probability equal to
unity) the value of a physical quantity, then  there exists an element of
physical reality corresponding to this physical quantity.'' 
  \cite{epr,jammer}. 
In Bohm's version \cite{bohm}, consider a pair of 
spin-$\frac{1}{2}$ particles in a singlet state
\begin{equation}
|\Psi>\,=\,\frac{1}{2}(|\uparrow_{1}\downarrow_{2}>-
|\downarrow_{1}\uparrow_{2}>) \label{bohm},
\end{equation}
 which is an eigenstate of $\hat{\sigma}_{1\hat{n}}\hat{\sigma}_{2\hat{n}}$,
 with $\hat{n}$ an arbitrary direction. The spin component of particle $2$
 may be predicted with certainty by measuring that of particle $1$ in the 
 same direction, and vice versa. Therefore every spin component of 
 each particle is an element of reality, and thus quantum mechanics is
 incomplete. EPR believed a complete theory is possible.
  von Neumann \cite{neumann}, Gleason \cite{gleason}, Jauch and Piron
 \cite{jp} gave proofs of impossibility of a hidden variable theory.
 Bell pointed out
 unreasonable postulates in these proofs by
   noting that noncommuting observables are measured in different
  experiments \cite{bell1}.
  Later he  proved  the Bell theorem that a local 
  hidden variable theory is inconsistent with quantum mechanics
  by deriving an inequality from the 
  hidden variables for EPR-Bohm setup,
  this equality is violated by quantum mechanics \cite{bell2}.
  On the other hand, Kochen and Specker (KS) also
  excluded the possibility
   of (noncontextual)
   hidden variables by proving that it is impossible to
   assign definite value to each of a set of   commuting 
   observables \cite{ks}. Recently Bell theorem without inequality
   received much interest. An earlier proof 
   was given  by  Heywood and Redhead \cite{heywood}
   with the aid of KS theorem. A few years ago
 Greenberger, Horn and Zeilinger gave a proof with EPR-Bell argument 
   for certain
 states of three or
   more spin-$\frac{1}{2}$ particles \cite{ghz},
   Mermin gave a simplified version \cite{mermin}. Hady gave proofs
   for two spin-$\frac{1}{2}$ particles
 \cite{hardy},  a streamlined version was given by Goldstein
  \cite{goldstein}. Nonlocality of
  a single photon was also claimed by Hardy\cite{hardy2}, while
   interpreted as multiparticle state by
  others \cite{vaidman2}. Among recent progress on KS theorem, 
  Peres gave a simple KS argument but relied on  
  singlet state \cite{peres},
   Mermin gave simple general KS arguments for two and  three
   spin-$\frac{1}{2}$ particles, 
   and found the relevance of the latter
   with  Bell  theorem \cite{mermin2}.  

Redhead extended EPR sufficient condition  
for the element of physical reality to: ``If we can predict with
 certainty, or at any rate with probability one, the result of measuring
 a physical quantity at time $t$, then at time $t$, there exists an 
 element of reality corresponding to this physical quantity and having a
 value equal to the predicted measurement result'' \cite{redhead}.
Redhead condition does not insist on  ``without in any way disturbing a 
 system'' though stresses ``at time $t$''.
 Actually every physical process happens at a certain time, so
 ``at time $t$'' is also implicit in EPR condition. In the 
 gedanken experiments of EPR and Bohm,
  the Hamiltonians are zero, therefore the elements of 
 reality do not change with time and thus the time is not explicit.
Therefore Redhead condition is weaker than EPR condition
while  the converse is stronger.
We will consider both. We also extend ``predict'' to ``infer'',
as done by Vaidman for Redhead condition \cite{vaidman}.

We will explain  that in all the above-mentioned
   work after EPR,  there is an implicit premise
   underlying the deduction leading
   to contradiction: the reality 
   is ``partial'', i.e., the elements of reality
   obey a ``sum rule'' and a 
   ``product rule''. However, recently
   Vaidman showed that the 
   ``product rule'' should be abandoned \cite{vaidman}.
   We  add that ``sum rule'' should also be abandoned; actually
   ``sum rule'' is a sufficient condition of ``product rule''.
Therefore all the arguments against hidden variables can be refuted
in a unified way.
In particular, since ``partiality'' of reality
is a premise in the proofs of Bell theorem,
  local realism
might be saved.

The notations are described.
The letter with a hat, like $\hat{A}$,
 represents an obsevable (physical quantity), as
well as the corresponding operator in quantum mechanics. The letter 
without a hat, like $A$,
 denotes the result of measuring $\hat{A}$.
The element of reality corresponding
to the observable $\hat{A}$ 
is denoted as $\{\hat{A}\}$, or $\{\hat{A}\}(\lambda)$, where 
$\lambda$ is the hidden variable.
Therefore if $A$ is obtained with probability equal to $1$,
 then   $\{\hat{A}\}(\lambda)\,=\,A$;
 if $A$ is  obtained with probability
 not equal to $1$, then it is unknown whether 
 $\{A\}$ exists.
 
First we birefly review Vaidman's results.
 Consider a quantum system which is 
prepared in a state $|\Psi_{1}>$
 at $t_{1}\,<\,t$, and is found in state  $|\Psi_{2}>$  at $t_{2}\,>\,t$.
The Hamiltonian is let to be zero for simplicity.
 Suppose $\hat{A}$ is measured at $t$.
If either $|\Psi_{1}>$
or  $|\Psi_{2}>$ is an eigenstate of $\hat{A}$, the outcome is certainly
the corresponding eigenvalue.
For a pre- and postselected
system it might be that the result of measuring $\hat{A}$ is certain,
i.e., with probability equal to $1$,
even though neither   $|\Psi_{1}>$
nor  $|\Psi_{2}>$  is an eigenstate of   $\hat{A}$. Suppose
 $\hat{B}$  can also  be measured with probability equal 
 to $1$.  Vaidman showed that the outcome of measuring $\hat{A}\hat{B}$ 
 may be uncertain, and may  be 
 certain but needs  not equal to the product of results
 of respective measurements
 of $\hat{A}$ and $\hat{B}$.
 Considering elements of reality defined by both prediction and 
 retrodiction,
 Vaidman said  that 
 Redhead condition 
  continues to hold
 with  ``predict''  changed  to ``infer''. 
 while the ``product 
rule'' should be abandoned. Using our notations, the result is as 
follows.
The existence of  $\{\hat{A}_{1}\}$ and $\{\hat{A}_{2}\}$ does not imply 
the existence of 
$\{\hat{A}_{1}\hat{A}_{2}\}=$$\{\hat{A}_{1}\}\{\hat{A}_{2}\}$. Even if
  $\{\hat{A}_{1}\hat{A}_{2}\}$ exists, it needs not equal 
  $\{\hat{A}_{1}\}\{\hat{A}_{2}\}$.
 Vaidman used his result to refute previous arguments against 
 Lorentz invariance \cite{hardy,clifton}.
The above result is a starting point of our work. However, we do not 
 agree on his interpretation of the 
 breakdown of ``product rule''	 as
 that  ``joint measurements of commuting operators in the considered
 situations invariably disturb each other'', therefore might
 be a manifestation 
 of nonlocality \cite{bendl}.
  We will come  back to this point finally.  
  
 Now we extend the beakdown of ``product rule'' to ``sum rule'' 
 which asserts that
the existence of  $\{\hat{A}_{1}\}$ and $\{\hat{A}_{2}\}$  implies 
the existence of 
$\{r_{1}\hat{A}_{1}+r_{2}\hat{A}_{2}\}$
$=r_{1}\{\hat{A}_{1}\}+r_{2}\{\hat{A}_{2}\}$,
where $r_{1}$ and $r_{2}$ are real numbers. 

As done by Vaidman, consider
EPR-Bohm setup.  At time $t_{1}$,
 $|\Psi_{1}\,
\,=|\Psi>$  as given by Eq. (\ref{bohm}).
At time $t_{2}$, $\hat{\sigma}_{1x}$ and  
$\hat{\sigma}_{2y}$  are
measured 
and certain results are obtained. Suppose the results are
$\sigma_{1x}\,=\,1$ and $\sigma_{2y}\,=\,1$, then $|\Psi_{2}>\,=\,
|\uparrow_{1x}\uparrow_{2y}>$.  
If at time $t$, $t_{1}\,<\,t\,<\,t_{2}$,
a measurement is performed on $\hat{\sigma}_{1y}$,
then the outcome that $\sigma_{1y}(t)\,=\,-\sigma_{2y}(t_{2})\,=-1$
is certain. If, instead, a measurement is performed on  $\hat{\sigma}_{2x}$,
the outcome is also certain:  
$\sigma_{2x}(t)\,=\,-\sigma_{1x}(t_{2})\,=-1$.
This   can be verifed 
 by the formula calculating probabilities for the results
of an intermediate measurement 
performed on a pre- and postselected system \cite{aharonov}:
\begin{equation}
p(A=a_{n})\,=\,\frac{|<\Psi_{2}|\hat{P}(A=a_{n})|\Psi_{1}>|^{2}}
{\sum_{k}|<\Psi_{2}|\hat{P}(A=a_{k})|\Psi_{1}>|^{2}}, \label{av}
\end{equation}
where $p(A=a_{n})$ is the probability for an intermediate measurement of
 $\hat{A}$ between $|\Psi_{1}>$ and $|\Psi_{2}>$ to yield $A\,=\,a_{n}$, 
$\hat{P}(A=a_{k})$ is the projection operator onto the subspace with 
eigenvalue $a_{k}$. However, Vaidman showed that
 $p(\sigma_{1y}\sigma_{2x}=1)=0$, violating  the ``product rule''.
 
Now we calculate   $p(\sigma_{1y}+\sigma_{2x}=-2)$. The relevant
 projection operators are
 $\hat{P}(\sigma_{1y}+\sigma_{2x}=2)$$=$$
 |\uparrow_{1y}\uparrow_{2x}><\uparrow_{1y}\uparrow_{2x}|$,
 $\hat{P}(\sigma_{1y}+\sigma_{2x}=0)$$=$$
 |\uparrow_{1y}\downarrow_{2x}><\uparrow_{1y}\downarrow_{2x}>$$
 +|\downarrow_{1y}\uparrow_{2x}><\downarrow_{1y}\uparrow_{2x}|$,
$\hat{P}(\sigma_{1y}+\sigma_{2x}=-2)$$=$$
 |\downarrow_{1y}\downarrow_{2x}><\downarrow_{1y}\downarrow_{2x}>$.
  Then Eq. (\ref{av}) yields $p(\sigma_{1y}+\sigma_{2x}=-2)\,=\,1/6$,
 $p(\sigma_{1y}+\sigma_{2x}=2)\,=\,1/6$, and
 $p(\sigma_{1y}+\sigma_{2x}=0)\,=\,2/3$. Therefore the ``sum rule'' is
 also violated; it does not follow that
 $\sigma_{1y}+\sigma_{2x}$ is  an element of 
 reality. It can be checked that both ``product rule'' and 
 ``sum rule'' are violated if $|\Psi_{2}>$ is any of the
  nine product states.
   Note that $\hat{\sigma}_{1y}+\hat{\sigma}_{2x}$ can be measured 
 using local interactions \cite{aharonov2}. 
 
From the definition of element of reality, the existence
of $\{\hat{A}\}$ implies the existence
of  $\{\hat{A}^{2}\}\,=\,\{\hat{A}\}^{2}$,
 since we can infer
the result of measuring $\hat{A}^{2}$ to be $\{\hat{A}\}^{2}$
  with probability equal to 
$1$ if $\{\hat{A}\}$ exists. Therefore it can be shown
as follows  that 
the validity of ``sum rule'' implies the validity of
``product rule'', and thus violation of the latter implies
violation of the former. Assuming the ``sum rule'', then
$\{\hat{A}_{1}\}^{2}+2\{\hat{A}_{1}\}\{\hat{A}_{2}\}+
\{\hat{A}_{2}\}^{2}$ $=$ 
$[\{\hat{A}_{1}\}+\{\hat{A}_{2}\}]^{2}$ $=$   
$\{\hat{A}_{1}+\hat{A}_{2}\}^{2}$ $=$ 
$\{(\hat{A}_{1}+\hat{A}_{2})^{2}\}$ $=$ 
$\{\hat{A}_{1}^{2}+2\hat{A}_{1}\hat{A}_{2}+\hat{A}_{2}^{2}\}$ $=$ 
$\{\hat{A}_{1}\}^{2}+2\{\hat{A}_{1}\hat{A}_{2}\}+
\{\hat{A}_{2}\}^{2}$, 
therefore the ``product rule'' that
$\{\hat{A}_{1}\hat{A}_{2}\}\,=\,\{\hat{A}_{1}\}\{\hat{A}_{2}\}$ 
is obtained. 
This deduction is similar to that
of Kochen and Specker for  the multiplicativity 
 from the additivity of  functions of an observable
 \cite{ks}.
 
Therefore, study on pre- and postselected quantum system
reminds us that  the reality is non-partial.
We call the  reality  ``partial'' if 
the ``sum rule'' and ``product 
rule'' are valid, i.e., the existence of elements of reality 
$\{\hat{A}_{1}\}$ 
and $\{\hat{A}_{2}\}$ implies the existence of elements 
$\{r_{1}\hat{A}_{1}+r_{2}\hat{A}_{2}\}$, which equals 
$r_{1}\{\hat{A}_{1}\}+r_{2}\{\hat{A}_{2}\}$, and  the existence of
$\{\hat{A}_{1}\hat{A}_{2}\}$, which equals
$\{\hat{A}_{1}\}\{\hat{A}_{2}\}$, 
where $r_{1}$ and $r_{2}$ are real numbers.

We use the term ``partial'' since 
according to Kochen and Specker a set of observable is called
a ``partial algebra'' 
if it is closed under the formation of ``partial operations''
of sums and products for pairwise
commuting observables in accordance with the 
following rules: if $\hat{A}_{1}\,=\,f_{1}(\hat{B})$  and
$\hat{A}_{2}\,=\,f_{2}(\hat{B})$, then $r_{1}\hat{A}_{1}+r_{2}\hat{A}_{2}$
$=$ $(r_{1}f_{1}+r_{2}f_{2})(\hat{B})$ and 
$\hat{A}_{1}\hat{A}_{2}\,=\,
(f_{1}f_{2})(\hat{B})$. 
They argued that a necessary condition for the existence of hidden 
variables is that the partial algebra be imbeddable in a commutative
 algebra (such as the algebra of all real-valued function on a phase space),
  and thus be preserved by the functions
  $f_{\hat{A}}(\lambda)\,=\,A$ corresponding to
  the observable $\hat{A}$ and its eigenvalue $A$, where $\lambda$
  is the hidden variable. 
  Therefore the existence of a homomorphism
 of the partial algebra of Hermitian operator  into the real set
 is necessary. Based on the nonexistence of such homomorphism, possibility 
 of (noncontextual) hidden  variables was excluded. This was demonstrated
 by considering the angular momentum operator equation
 $\hat{J}^{2}=\hat{J}_{x}^{2}+\hat{J}_{y}^{2}+\hat{J}_{z}^{2}$  in
 $j=1$ state. Since $\hat{J}_{x}^{2}$, $\hat{J}_{y}^{2}$, 
 $\hat{J}_{z}^{2}$ commute 
 among themselves as well as with $\hat{J}^{2}$ which yields $2$
 and are therefore simultaneously measurable, always yielding
 one $0$ and two $1$, a hidden variable theory should assign 
 to each direction a definite value of  the component of
 $\hat{J}^{2}$.  It was shown that no such assignment exists.
 In the argument of Heywood and Redhead,  EPR argument for a pair of
 spin-$1$ particles with total spin zero yields the conclusion that
 there exists definite value of $0$ or $1$ for each component of
 $\hat{J}^{2}$, contradicting KS theorem.
 Now we know that in the pre- and postselected quantum system,
 the partial algebra is not preserved. 
 Though $\hat{A}_{1}$ and $\hat{A}_{2}$ commute and can be 
 measured with probability $1$,
 a partial operation of them 
 might not. Even if  it can be measured with
 probability $1$, the result might not be the partial operation
 of the results of $A_{1}$ and $A_{2}$. It implies that the 
  reality  is not partial, and thus the anticipated 
   hidden variable theory 
  giving the functions $f_{\hat{A}}(\lambda)\,=\,A$  does not
  embed partial algebraic structure. In the example of $j=1$ state,
  although an anticipated
   hidden variable theory gives definite value (an 
  element of reality) of either $0$ or $1$
  to each  component of $\hat{J}^{2}$,
   and definite value of $2$ to $J^{2}$, there
  is no reason to require these elements of reality to
  obey the same equation of the operators, i.e. the ``sum rule'' is
  violated.  In the recent simplified forms of Kochen-Specker argument
  using two or three spin-$\frac{1}{2}$ particles by
  Peres and by Mermin,
  instead of ``sum rule'', ``product rule''  was used.

  Bell pointed out the irrelevance of von Neuman's impossibility proof
  by rejecting the postulate 
 of additivity of expectation value.
    Gleason theorem  reduced the additivity postulate to
    be only of commuting observables.
    But Bell continued to
    show that in the  corollary excluding hidden variables,
  as well as in the  version of Jauch and Piron, it was still assumed,
   in a tacit way, that the measuring result of an observable
    is independent of 
   measurement made on other observables not commuting with the former.
    Now we can see that in addition to Bell's arguments,
 the  postulatate of additivity of
 expectation value even only for commuting observable  is 
 unrealistic given the  non-partiality of reality.

  In deriving  Bell theorem in the  
  form of  inequality for EPR-Bohm setup, 
  Bell studied the hidden variable theoretic expectation value 
$P(\hat{n}_{1},\hat{n}_{2})$ $=\,\int d\lambda\rho(\lambda)
\{\hat{\sigma}_{1\hat{n}_{1}}\}(\lambda)
\{\hat{\sigma}_{2\hat{n}_{2}}\}
(\lambda)$ supposed to correspond to quantum mechanical expectation
value $<\hat{\sigma}_{1\hat{n}_{1}}\hat{\sigma}_{2\hat{n}_{2}}>$.
Then as a special case, for the perfect correlation, 
there must be
 $\{\sigma_{1\hat{n}}\}(\lambda)=-\{\sigma_{2\hat{n}}\}(\lambda)$
 so that $P(\hat{n},\hat{n})\,=\,-1$ as required by quantum mechanical 
 result.
 In his derivation, the ``product rule'' was used.
  It was
 implicitly assumed that corresponding
 to   $\hat{\sigma}_{1\hat{n}_{1}}\hat{\sigma}_{2\hat{n}_{2}}$
 there must exists element of reality 
$\{\hat{\sigma}_{1\hat{n}_{1}}\hat{\sigma}_{2\hat{n}_{2}}\}$ and it
equals to
$\{\hat{\sigma}_{1\hat{n}_{1}}\}\{\hat{\sigma}_{2\hat{n}_{2}}\}$ . 
 Through
 EPR argument, one can be convinced that there exist
elements of reality corresponding to 
  $\hat{\sigma}_{1\hat{n}_{1}}$ and $\hat{\sigma}_{2\hat{n}_{2}}$ for
arbitrary $\hat{n}_{1}$ and $\hat{n}_{2}$, but it does not follow
that there is an element of reality corresponding to their product,
let alone this element of reality is just equal to
 $\{\hat{\sigma}_{1\hat{n}_{1}}\}\{\hat{\sigma}_{2\hat{n}_{2}}\}$.
 In another word, $P(\hat{n}_{1},\hat{n}_{2})$ given 
 above does not correspond to 
 $<\hat{\sigma}_{1\hat{n}_{1}}\hat{\sigma}_{2\hat{n}_{2}}>$ if  the 
 ``product rule'' is abandoned.

The  perfect correlation needs more analyses.
In addition to Bell's method  obtaining 
 $\{\hat{\sigma}_{1\hat{n}}\}\,=\,-\{\hat{\sigma}_{2\hat{n}}\}$, 
 there might be
 the following reasoning without considering the general
 $P(\hat{n}_{1},\hat{n}_{2})$.
 Since the singlet state (\ref{bohm}) is an eigenstate of  
  $\hat{\sigma}_{1\hat{n}}\hat{\sigma}_{2\hat{n}}$, it is known that
    $\{\hat{\sigma}_{1\hat{n}}\hat{\sigma}_{2\hat{n}}\}$=$
    \sigma_{1\hat{n}}\sigma_{2\hat{n}}=-1$ is an
  element of reality according to Redhead condition (this might be
  doubtful if EPR's ``without disturbing'' is  insisted). Then
  two alternative ways might follow. 
  (a) Use the ``product rule'' to obtain 
  $\{\hat{\sigma}_{1\hat{n}}\}\{\hat{\sigma}_{2\hat{n}}\}\,=$ 
  $\{\sigma_{1\hat{n}}\sigma_{2\hat{n}}\}\,=\,-1$.
  (b) It is convinced by EPR argument  that
  $\{\sigma_{1\hat{n}}\}$ and $\{\sigma_{2\hat{n}}\}$
  exist. Hence in the identity
 $\sigma_{1\hat{n}}\sigma_{2\hat{n}}\,=\,-1$,
   one just replace 
   $\sigma_{1\hat{n}}$ with $\{\sigma_{1\hat{n}}\}$,
   and $\sigma_{2\hat{n}}$ with $\{\sigma_{2\hat{n}}\}$,
   or considering  $\{\hat{\sigma}_{1\hat{n}}\}$
($\{\hat{\sigma}_{2\hat{n}}\}$) is inferred by 
$\sigma_{2\hat{n}}$ ($\sigma_{1\hat{n}}$),
   replace 
   $\sigma_{1\hat{n}}$ with $-\{\hat{\sigma}_{2\hat{n}}\}$
   and $\sigma_{2\hat{n}}$ with $-\{\hat{\sigma}_{1\hat{n}}\}$.
But we have to say  that this is a circular reasoning:
 $\{\hat{\sigma}_{1\hat{n}}\}$
 = $\sigma_{1\hat{n}}$ = $-\sigma_{2\hat{n}}$ or
$\{\hat{\sigma}_{2\hat{n}}\}$
 = $\sigma_{2\hat{n}}$ = $-\sigma_{1\hat{n}}$ is just deducted from
 $\sigma_{1\hat{n}}\sigma_{2\hat{n}}\,=\,-1$, one cannot 
 put the result back to get more result;
 by doing this, it means that it has been certain that the measuring 
 result on another particle, by which the element of reality for
 one particle is inferred,
  is also an element of reality in the meantime ,
 but this is just the anticipated result of doing this.
  From physics viewpoint,
 particle $2$ must be disturbed in order that
   $\{\hat{\sigma}_{1\hat{n}}\}$
is inferred, and vice versa.
 The probability
of  inferring  $\{\hat{\sigma}_{1\hat{n}}\}$,
  which is equal to $1$, is actually the conditional probability that 
 $\{\hat{\sigma}_{1\hat{n}}\}\,=\,-\sigma_{2\hat{n}}$;
  because there
 is no causal connection between these two particles the 
 consitional probability is just equal to the probability of inferring
  $\{\hat{\sigma}_{1\hat{n}}\}$. But in the meantime,
 the value of $\sigma_{2\hat{n}}$  cannot be obtained
 with
 certainty.
 Therefore the 
perfect correlation 
is not between elements of reality but between the 
 element of reality of one particle 
and  the measuring result of another particle, 
which is  obtained with disturbance and with a  
probability not equal to $1$.  In conclusion,
$\{\hat{\sigma}_{1\hat{n}}\hat{\sigma}_{2\hat{n}}\}\,=\,-1$ 
cannot be 
changed to 
$\{\hat{\sigma}_{1\hat{n}}\}\{\hat{\sigma}_{2\hat{n}}\}\,=\,-1$. 

In GHZ-Mermin approaches to Bell theorem without inequality,
contradiction was found  by dealing with  
perfect correlation.
  GHZ's original method is just similar to Bell's. Consider, e.g,
  $|\Psi>\,=\,$$\frac{1}{\sqrt{2}}(|\uparrow_{1}\uparrow_{2}
  \downarrow_{3}
    \downarrow_{4}>-|\downarrow_{1}\downarrow_{2}\uparrow_{3}
    \uparrow_{4}>$.
    Corresponding to
  $<\hat{\sigma}_{1\phi_{1}}\hat{\sigma}_{2\phi_{2}}
    \hat{\sigma}_{2\phi_{2}}\hat{\sigma}_{2\phi_{2}}>$, 
      where spins are assumed to be co-planar for simplicity,	the
``product rule''
    is assumed to obtain $P\,=\,$$\int d\lambda\rho(\lambda)
    \{\hat{\sigma}_{1\phi_{1}}\}(\lambda)
    \{\hat{\sigma}_{2\phi_{2}}\}(\lambda) 
    \{\hat{\sigma}_{3\phi_{3}}\}(\lambda)
    \{\hat{\sigma}_{4\phi_{4}}\}(\lambda)$,
 though this expression was not written out explicitly.
      Then  from quantum mechanical result for perfect correlation,
    several equations identifying the integrand
        to $1$ or $-1$  are given and lead to 
contradiction.  
    In Mermin's version, consider, e.g, $|\Psi>\,=$
    $\frac{1}{\sqrt{2}}(|\uparrow_{1}\uparrow_{2}\uparrow_{3}>-
    |\downarrow_{1}\downarrow_{2}\downarrow_{3}>)$,
     which is a simultaneous
    eigenstate of 
    $\hat{\sigma}_{1x}\hat{\sigma}_{2y}\hat{\sigma}_{3y}$,
$\hat{\sigma}_{1y}\hat{\sigma}_{2x}\hat{\sigma}_{3y}$,
$\hat{\sigma}_{1y}\hat{\sigma}_{2y}\hat{\sigma}_{3x}$, and
$\hat{\sigma}_{1x}\hat{\sigma}_{2x}\hat{\sigma}_{3x}$,
with   eigenvalues $1$, $1$, $1$, and $-1$, respectively.
In the four identities, each spin operator is replaced by the corresponding 
element of reality, then inconsistency is obtained. This
 reasoning  is
just the circular one analyzed above.
 
 Hardy-Goldstein approaches were also based on ``product rule''.
 It suffices to anylyze the streamlined version by Goldstein.
 Consider 
$|\Psi>\,=\,a|v_{1}>|v_{2}>+b_{1}|u_{1}>|v_{2}>+b_{2}|v_{1}>|u_{2}>$, where
$ab_{1}b_{2}\neq 0$, $|u_{i}>$ and $|v_{i}>$ are basis of particle $i$. Let
$\hat{U}_{1}=|u_{1}><u_{1}|$ etc., and
$|w_{i}>\propto a|v_{i}>+b_{i}|u_{i}>$. One has: (i) $U_{1}U_{2}=0$.
(ii) $U_{1}=0$ implies $W_{2}=1$ while  $U_{2}=0$ implies $W_{1}=1$.
Therefore assumption of local hidden variables contradict 
(iii) $W_{1}=W_{2}=0$ with nonvanishing probability.
In this deduction, perfect correlation
(i) implies $\{U_{1}U_{2}\}=0$, but
 $\{U_{1}\}\{U_{2}\}=0$ does not follow if the ``product rule''
 is abandoned. Then the implication (ii) from $|\Psi>$ 
 cannot be made; one should consider the collapsed state as follows.
  Suppose $U_{2}$ is measured, if $U_{2}=0$,
 the state collapses to be $|w_{1}>|v_{2}>$, indeed
 $W_{1}=1$; but if $U_{2}=1$, the state becomes
 $c|v_{1}>|u_{2}>$:  now it is certain $U_{1}=0$,
 but there are both non-zero 
 probabilities for $W_{2}$ to be $1$ and $0$, in the
 meantime $W_{1}$ may also be  $0$ or $1$. So there is
 no contradiction with (iii). Only if  $\{U_{1}\}\{U_{2}\}=0$
  can Goldstein's resoning  be valid.

  Both Bell Theorem and the simplified form of Kochen-Specker arguments
of Peres and Mermin  use ``product rule'', 
which is applied to arbitrary observables
 for the latter, while only to causally disconnected ones
 for the  
former. It is  also found that the ``product rule'' is related to the 
``sum rule'', which was
used in original KS theorem.
This appears to be 
the origin of the relation  between simple forms of the two Theorems 
exposed by Mermin in GHZ state. 

Finally, we note that  non-partiality is not identical with 
either contextual or 
nonlocality.
 Non-partiality means that the element of reality  corresponding to
 the sum or product of two observables is not the same partial
 operation of the 
  elements of reality corresponding to the two observables,
 while contextuality concerns the relation between the two
independent observables; referring to that the measuring result of one
 observable depends on the simultaneous  measurement made on 
  the other    
 observable, even though they are compatible. Similarly, nonlocality
 also concerns the relation between two independent observables;
 in this case they are causally disconnected. Although contextuality or nonlocality
 gives rise to non-partiality (this is  why non-local hidden variable
 theory, such as that of Bohm \cite{bohm2}, can work), 
 non-partiality might be
 noncontextual and local. In addition to logical reasoning, a
  support to this possibility comes from that 
  the partial operation of operators of causally disconnected particles
  might be 
  measured locally \cite{aharonov2}.
  Further clarification on this point is deserved.

  To summarize, study on pre- and postselected quantum systems 
  indicates that both ``product rule'' and ``sum rule'' for
  the elements of reality should be abandoned. We expose that 
  all arguments 
   against 
  hidden variables were based on the assumption of  partiality of reality,
  therefore can be refuted unifiedly by non-partial realism, which might
  thus save local realism. At least, we have reached a unified
  understanding of  Bell theorem  and KS theorem.

\end{document}